\documentclass[FBSedit,FBSmath,ecsub]{FBSsuppl}
\usepackage{amsfonts}
\usepackage{amssymb}
\usepackage{epsfig}

\newcommand{\gl}{\lambda}

\newcommand{\gL}{\Lambda}

\newcommand{\be}{\begin{equation}}
\newcommand{\ee}{\end{equation}}
\title{Three quark clusters in hot and dense matter}

\author{S.  Mattiello$^a$\thanks{{\it E-mail address:}
    stefano.mattiello@physik.uni-rostock.de}, M. Beyer$^a$
, T. Frederico$^b$, H. J. Weber$^c$}

\institute{$^a$Fachbereich Physik, Universit\"at Rostock, D-18051
  Rostock, Germany\\$^b$Dep. de F\'\i sica, ITA, CTA, 12.228-900 S\~ao Jos\'e dos Campos, Brazil\\
  $^c$Dept. of Physics, University of Virginia, Charlottesville, VA
  22904, U.S.A.}

\begin{document}

\maketitle

\begin{figure}[t]
\begin{minipage}[t]{0.48\textwidth}
\epsfig{figure=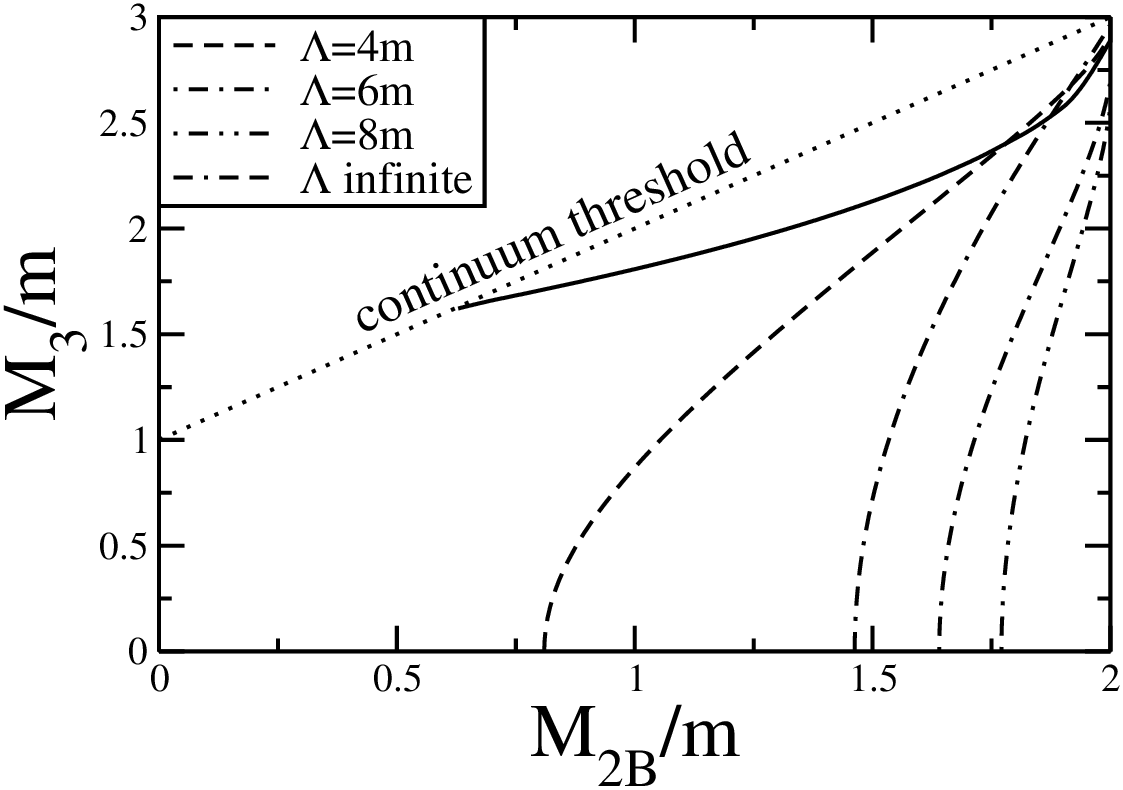,width=\textwidth}
\caption{\label{m2-m3} $M_3$ as function of the mass of the "diquark" $M_{2B}$.}
\end{minipage}
\hfill
\begin{minipage}[t]{0.48\textwidth}
\epsfig{figure=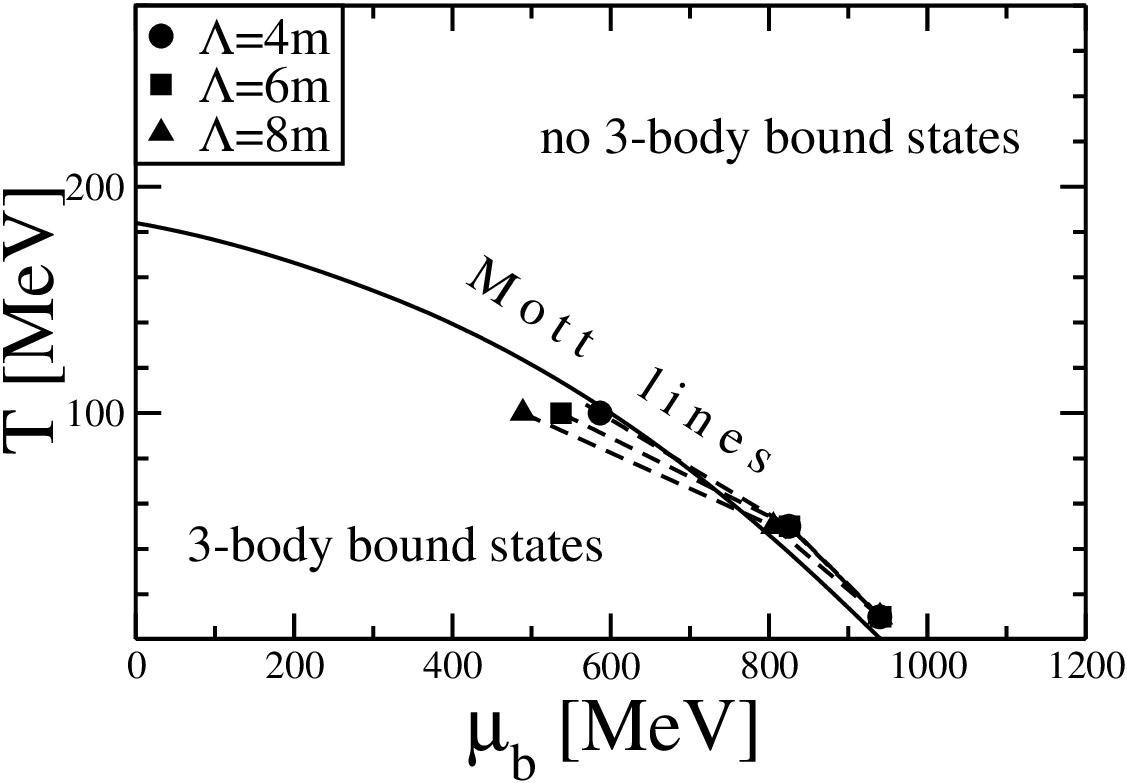,width=\textwidth}
\caption{\label{phase} Phase-diagram of QCD. See text for explanation.}
\end{minipage}
\end{figure}

The phase diagram of quantum chromo\-dy\-na\-mics (QCD) shows a rich structure as lattice simulations and model calculations indicate~\cite{SC}.
One very interesting aspect is the transition from quarks to nucleons
as relevant degrees of freedom (Mott transition~\cite{FBS01,PLB01}). This is related to
the 
transition between the hadronic
matter and the quark-gluon plasma phase.
Another particularly interesting possibility is quark pairing that leads to
the color superconductivity analogous to Cooper pairing.
An aspect that has hardly been investigated in this context is the
appearance of three-quark correlations that, in the vicinity of the phase transition, should play an important role and  might as
well influence quark pairing. Here we address the issue of three-quark
clusters in a medium of finite temperatures and densities. This necessarily involves relativity as well
as effects of the medium.
To implement relativity we use the light front
approach and utilize a Green function formalism.
We consider first the isolated case and then medium effects. Using a zero-range
interaction, the two-body propagator is $t=(i\gl^{-1}-B(M_2))^{-1}$, where $B(M_2)$ denotes the loop momentum integral~\cite{Tobias}. The logarithmic divergence appearing in
the integral can be eliminated by introducing  an invariant cut-off
$\gL$. The requirement is that $\gL$ is smaller
than the intermediate mass of the virtual two-quark subsystem,
i.e. $M_{20}^2<\gL^2$ and hence $t\rightarrow t_\gL$. We introduce a similar regularisation for the
three-body equation ($M_{30}^2<\gL^2$) that leads to an equation for
the vertex function $\Gamma_\gL$ that depends parametrically on the cut-off $\gL$,
\begin{displaymath}
\Gamma_\gL(M_3;y,\vec q_\perp) = \frac{i}{(2\pi)^3}\ t_\gL(M_2)\!
\int_{x_{\mathrm{min}}}^{1-y} \!
\frac{dx}{x(1-y-x)}\!
\end{displaymath}
\begin{displaymath}
\int\! d^2k_\perp
\frac{\theta\left(M_{03}^2-\gL^2\right)}
{M^2_3 -M_{03}^2}
\Gamma_\gL(M_3;x,\vec k_\perp).
\end{displaymath}
In Fig.~\ref{m2-m3} we show the mass of the three-quark cluster as
function of the diquark mass for different cut-offs in units of the
quark mass. Note that $M_3\rightarrow0$ for finite diquark masses which
is the relativistic analog of the Thomas collapse~\cite{Thomas}. The
results using the previous regularisation~\cite{Tobias} are shown as solid line.\\
\indent
To investigate medium effects we fit the parameters $(\gL,\gl)$ to
reproduce the proton mass. The three-body equation systematically includes the dominant medium effects due to Pauli
blocking and self energy corrections, $m\rightarrow m(T,\mu)$. For a
given temperature and a given choice of the fit parameters,  the
three-quark system becomes more weakly bound, as the baryonic chemical
potential $\mu_{\mathrm{b}}$ increases. The continuum is
reached for  binding energy $B_3=0$ (Mott transition).
The Mott lines, i.e. the values of $T$ and $\mu_{\mathrm{b}}$ where this transition
occurs are given in Fig.~\ref{phase}.
We show our previous result (solid line) along with the ones using the
invariant cut-off. 
From other models investigating $q\bar{q}$ states the confinement-deconfinement phase
transition has been estimated as well. We find a similar qualitative
behavior for the three-body Mott lines. A closer comparison indicates
differences at larger densities that are currently investigated.
The next aim is to investigate the phase transition between the color
superconducting phase and the quark matter considering the influence
of three-quark-correlations.

\end{document}